\documentclass{article}
\usepackage{amsmath}
\usepackage{amssymb}
\usepackage{amsthm}
\usepackage{natbib}
\usepackage{epsfig}
\usepackage[tight]{subfigure}

\title{Reduced form models of bond portfolios}

\author{Matti Koivu\thanks{Finnish Financial Supervisory Authority, Market and Operational Risk Division, \texttt{matti.koivu@bof.fi}. The views presented in this paper are those of the authors' and are not necessarily shared by the Finnish Financial Supervisory Authority} \and Teemu Pennanen\thanks{Department of Mathematics and Systems Analysis, Aalto University, \texttt{teemu.pennanen@tkk.fi}, corresponding author}}

\begin{document}

\maketitle

\abstract{We derive simple return models for several classes of bond portfolios. With only one or two risk factors our models are able to explain most of the return variations in portfolios of fixed rate government bonds, inflation linked government bonds and investment grade corporate bonds. The underlying risk factors have natural interpretations which make the models well suited for risk management and portfolio design.}

\section{Introduction}

Bond portfolios are subject to a large number of risk factors. Even in the case of default-free fixed rate bonds, the portfolio return depends in general on the whole yield curve which is a high-dimensional object. Portfolio managers often describe returns in terms of {\em yield to maturity}, which provides a {\em one-dimensional} approximation of the expected return on a given portfolio. Such simple descriptions are easy to understand and to model, which makes them useful in risk management and asset allocation. 

This paper derives low-dimensional return formulas for several classes of bond portfolios, including fixed rate government bonds, index linked bonds and defaultable corporate bonds. The underlying risk factors in our models have natural interpretations which facilitates the assessment and modeling of the risks. For example, the returns on inflation linked bonds are quite accurately described in terms of the yield and the underlying consumer price index. Returns on corporate bonds, in turn, are well approximated by the yield and the yield spread between corporate and government bonds. With only one- or two risk factors, our models are able to describe consistently over 97\% of monthly return variations on government, inflation linked and corporate bonds over the past decades including the recent financial crisis.

Our models are based on low-order Taylor-approximations of the logarithmic price of the portfolio. This is analogous to \cite{cj96} where a Taylor-approximation of the price with respect to yield and time was found to give good approximations of return in the case of fixed-rate government bonds. Our model goes one step further by developing Taylor-approximations with respect to time, the yield as well as the outstanding coupon/principal payments. Variations in the outstanding payments are an essential return component e.g.\ in the case of index linked or corporate bonds. For inflation linked bonds, it amounts upto 10\% increase in the explained historical return variations. Even in the case of default free fixed-rate bonds, our models improve on earlier ones. Compared to return formulas obtained by low-order approximations of the price itself, the logarithmic approach simplifies the formulas and improves the accuracy of the approximations.

\section{Returns on bond portfolios}\label{sec:ret}

Consider a bond portfolio whose coupon and principal payments are, as they are received, reinvested in bonds at current market prices. The portfolio's {\em total return index} $P_t$ at time $t$ then equals the market value of the fund's assets at all times. The {\em yield to maturity}, $Y_t$ of the portfolio is defined through the equation
\begin{equation}\label{y}
P_t = \sum_{n=1}^N e^{-Y_t(t_n-t)}C_{t,n},
\end{equation}
where $C_{t,n}$ denotes the fund's outstanding aggregate coupon and principal payments payable at time $t_n>t$. Whereas the default-free yield curve can be used to price an arbitrary portfolio of default free bonds, the yield to maturity is defined for a particular portfolio of (default-free or not) bonds. The yield to maturity of a portfolio is sometimes called the ``internal rate of return''. The number $N$ of payment dates $t_n$ in \eqref{y} can be allowed to be infinite but in that case our subsequent analysis requires that the outstanding payments $C_{t,n}$ be bounded in $n$. An infinite $N$ may be useful when modeling perpetual bonds.

The definition of the outstanding payments is crucial in the definition of the yield. In the case of fixed rate bonds, there is little ambiguity: the outstanding payments are what the issuer has promised to deliver even though in the presence of default risk, the actual amount that will be received at time $t_n$ may be less than $C_{t,n}$. Default risk is reflected as a lower price and thus a higher yield. In the case of index linked bonds, the coupon payments are uncertain by definition and the outstanding payments in \eqref{y} need to be forecast somehow. The simplest option is to ``freeze'' the underlying index to its current value when estimating $C_{t,n}$  at time $t$. In the case of inflation linked bonds, the corresponding yield is known as the {\em real yield}. For now, it suffices to assume that the yield $Y_t$ is defined according to \eqref{y} whatever the definition of outstanding payments may be.

Equation \eqref{y} expresses the bond price as a function of time $t$, the yield $Y_t$ and the sequence $C_{t}=(C_{t,n})_{t=1}^N$ of payments outstanding at time $t$. Indeed, we have $P_t=P(t,Y_t,C_t)$, where $P$ is the function defined for each $t$, $Y$ and $C=(C_n)_{n=1}^N$ by
\[
P(t,Y,C) = \sum_{n=1}^N e^{-Y(t_n-t)}C_n.
\]
Similarly, the log-return on the portfolio over a holding period $[t,s]$ can be expressed as
\[
\Delta \ln P \approx \ln P(s,Y_s,C_s) - \ln P(t,Y_t,C_t).
\]
The approximation is exact if there are no payments during the holding period, i.e.\ if $t_1\ge s$. If $s>t_n$, the $n$th payment has been collected and reinvested in the portfolio. An error may result if the reinvested payment appreciates during $[t_n,s]$ at a rate different from $Y_s$. If the payments during $[t,s]$ amount to a small fraction of the all outstanding payments (if length of the holding period is small compared to the maturity of the bonds in the portfolio), the resulting error could be expected to be small.

Our reduced form models are based on Taylor-approximations of the log-return with respect to time, the yield and the outstanding payments. The first order approximation can be written as
\begin{equation}\label{ret0}
\Delta \ln P \approx\frac{1}{P_t}\frac{\partial P}{\partial t}(t,Y_t,C_t)\Delta t + \frac{1}{P_t}\frac{\partial P}{\partial Y}(t,Y_t,C_t)\Delta Y + \sum_{n=1}^N\frac{1}{P_t}\frac{\partial P}{\partial C_{n}}(t,Y_t,C_t)\Delta C_n
\end{equation}
where $\Delta Y=Y_{s}-Y_t$ and $\Delta C_{n}=C_{s,n}-C_{t,n}$. The first two terms are quite familiar. Indeed, we have 
\[
\frac{1}{P_t}\frac{\partial P}{\partial t}(t,Y_t,C_t) = Y_t\quad\text{and}\quad \frac{1}{P_t}\frac{\partial P}{\partial Y}(t,Y_t,C_t) = -D_t,
\]
where 
\[
D_t = \frac{1}{P_t}\sum_{n=1}^N (t_n-t)e^{-Y_t(t_n-t)}C_{t,n} 
\]
is the {\em Macaulay duration} of the portfolio at time $t$. The first term of \eqref{ret0} represents the {\em pull to par} effect, which means that bonds tend to appreciate when the maturity is approached. The second term gives the price sensitivity with respect to the yield. The effects of time and yield changes on returns off fixed-rate government bonds have been studied e.g.\ in \cite{cj96} who developed a Taylor series approximation for the price itself (as opposed to its logarithm). 

The last term in \eqref{ret0} is the return component arising from changes in outstanding payments during the holding period $[s,t]$. Because of its multivariate character, the last term is difficult to model in general. However, when there are no portfolio updates during the holding period, there is often a single risk factor $K_s$ such that
\begin{equation}\label{k}
\Delta C_{n}\approx K_sC_{t,n}
\end{equation}
for all $n=1,\ldots,N$. For default-free fixed coupon bonds, $\Delta C_{t,n}=0$ so \eqref{k}  holds trivially with $K_s=0$. In the case of index linked bonds, on the other hand, $K_s$ equals the change in the underlying index over the holding period $[t,s]$; see Section~\ref{sec:il} below. When \eqref{k} holds, the last term in \eqref{ret0} reduces to
\begin{align*}
\sum_{n=1}^N\frac{1}{P_t}\frac{\partial P}{\partial C_{n}}(t,Y_t,C_t)\Delta C_{t,n} &\approx K_s\frac{1}{P_t}\sum_{n=1}^Ne^{-Y_t(t_n-t)}C_{t,n} = K_s
\end{align*}
and \eqref{ret0} becomes
\begin{equation}\label{ret1}
\Delta\ln P \approx Y_t\Delta t - D_t\Delta Y + K_s. 
\end{equation}
The portfolio return over the holding period $[t,s]$ is then approximated by
\[
\frac{\Delta P}{P_{t}} \approx \exp\left(Y_t\Delta t - D_t\Delta Y_t + K_s\right)-1,
\]
where the right side is bounded from below by $-1$ just like actual portfolio return we are trying to approximate. This is why we linearized the logarithm of the price instead of the price itself as is often done. 

The logarithmic approach is supported also by the second order analysis. Denoting the convexity by
\[
C_t = \frac{1}{P}\sum_{n=1}^N(t_n-t)^2e^{-Y(t_n-t)}C_{t,n},
\]
we get, by straightforward differentiation, that
\[
\frac{\partial^2\ln P}{\partial t\partial Y} = 1,\quad \frac{\partial^2\ln P}{\partial Y^2}= C_t-D_t^2,\quad \frac{\partial^2\ln P}{\partial Y\partial C_n} = \frac{e^{-Y_t(t_n-t)}}{P}(D_t-t_n+t)
\]
while all other second order derivatives of the logarithmic price are zero. If \eqref{k} holds, the second order approximation of the log-return reduces (after simple algebraic manipulations) to
\begin{equation}\label{ret2nd}
\Delta\ln P \approx Y_{s}\Delta t - D_t\Delta Y + K_s + \frac{1}{2}(C_t-D_t^2)\Delta Y^2
\end{equation}
This differs from the first order formula \eqref{ret1} only by the addition of the term quadratic in $\Delta Y$ and in that, in the first term, $Y_t$ has been replaced by $Y_s$. This should be compared with the more complicated return formula of \cite{cj96} who studied fixed-rate bonds through a second order approximation of the price itself (instead of its logarithm). The reduction of the second order terms in the quadratic approximation of the log return suggests that the logarithm of the price is well approximated already by the first order terms. 




\section{Fixed rate government bonds}\label{sec:gov}

In the case of fixed rate default-free bonds, we have $\Delta C_{t,n}=0$ so that \eqref{k} holds with $K_s=0$ and \eqref{ret2nd} reduces to 
\begin{equation}\label{eqgov}
\Delta\ln P \approx Y_s\Delta t - D_t\Delta Y + \frac{1}{2}(C_t-D_t^2)\Delta Y^2.
\end{equation}
This is similar to the model studied e.g.\ in \cite{ilm92} but there the time component was ignored. \cite{cj96} incorporated the time component but their model, like that of \cite{ilm92}, was based on a Taylor-approximation of the price instead of its logarithm. 

\subsection{Empirical results}

We study the accuracy of the above models in explaining monthly returns on fixed rate government bonds. Our dataset covers end of month observations of the Barclays' market capitalisation weighted total return indices, durations and yields for France, Germany, Italy, United Kingdom, United States and the Euro area\footnote{Further information is available online at https://ecommerce.barcap.com/indices/index.dxml}. The length of the time series for each country is given at the bottom of Table \ref{tab:resfixgov}. 

We fit the following two models to the data
\begin{center}
\begin{tabular}{ll}
Model 1:	& $\Delta\ln P_s = c + Y_s\Delta t - D_t\Delta Y_s + \epsilon_s$, \\
Model 2:	& $\Delta\ln P_s = c + Y_s\Delta t - D_t\Delta Y_s + \gamma\Delta Y_t^2 + \epsilon_s$.
\end{tabular}
\end{center}
The parameters $c$ and $\gamma$ are estimated by ordinary least squares. The regression statistics provide us with diagnostic tools to evaluate the performance of the proposed models. By comparing the fits of the two models we can assess the significance of the quadratic term in explaining the returns.

Table~\ref{tab:resfixgov} displays the estimation results for the two model specifications. Model 1 already provides an almost perfect fit to the total return data with $R^2$ values ranging from 99.6\% to 99.8\%. The model fit has been consistent over time and the residuals have remained marginal even during the recent financial crisis. The $R^2$ values being close to 100\% there is not much room for improvement when adding the second order term. The estimated coefficients $\gamma$ in Model~2 are generally significant at a 5\% level but, consistently with previous research \citep{cj96}, the improvements in the $R^2$ statistics are marginal, less than $0.02\%$ points in all the studied markets. The Partial-$R^2$ statistic is defined as the $R^2$-statistic obtained by regressing the residual of Model~1 with the quadratic term. The quadratic term of Model 2 explains less than 5.11\% of the residual variance of Model~1. The estimated constant $c$ in Model~1 deviates from zero at 5\% significance level, but with varying signs, This may indicate that the model has not entirely captured all the systematic return components. The addition of the quadratic term in Model~2 mitigates this effect to some extent.

\begin{table}[ht!]
\begin{center}
\caption{Estimation results for fixed rate government bonds.}\label{tab:resfixgov}
{\small
\begin{tabular}{r|cccccc}
\hline
 &FRA	&GER &IT	&UK	&US	&EURO \\
\hline
\textbf{{Model 1}} &&&&&&\\
$100*c$	&0.0147	&0.0083	&0.0084	&-0.0195	&-0.0202	&0.015\\
				&(3.5842)	&(1.7215)	&(1.9339)	&(-4.8403)	&(-3.0658)	&(4.1467)\\
$R^2$		&99.76\%	&99.64\%	&99.68\%	&99.84\%	&99.71\%	&99.79\%\\
\textbf{{Model 2}} &&&&&&\\
$100*c$	&0.007	&0.0007	&-0.0001	&-0.0256	&-0.0277	&0.01\\
	&(1.4255)	&(0.1186)	&(-0.0178)	&(-5.7509)	&(-3.6689)	&(2.2687)\\
$\gamma$	&24.4889	&24.1545	&32.5155	&4.8396	&11.1262	&17.031\\
	&(2.7949)	&(2.2625)	&(2.5933)	&(3.0699)	&(1.9703)	&(1.9131)\\
$R^2$	&99.77\%	&99.65\%	&99.70\%	&99.85\%	&99.72\%	&99.79\%\\
Partial-$R^2$	&5.11\%	&3.41\%	&4.43\%	&2.63\%	&2.84\%	&2.46\% \\
\hline	
Data start	&1997-12	&1997-12	&1997-12 &1980-12	&1998-12		&1997-12\\
Data end 	&2010-03	&2010-03	&2010-03 &2010-03	&2010-03		&2010-03\\
\hline
\multicolumn{7}{l}{ {\it{Notes.}} The table contains estimation results for the models:} \\
\multicolumn{7}{l}{ Model 1: $\Delta\ln P_s = c + Y_s\Delta t - D_t\Delta Y_s + \epsilon_s$,}   \\
\multicolumn{7}{l}{ Model 2: $\Delta\ln P_s = c + Y_s\Delta t - D_t\Delta Y_s + \gamma\Delta Y_t^2 + \epsilon_s$,}\\
\multicolumn{7}{l}{ Numbers in parentheses are the $t$-statistics for the estimated coefficients.}	
\end{tabular}}	
\end{center}
\end{table}

Our results are consistent with \cite{ilm92} who finds that the explanatory power of duration has increased over time. In his empirical study, \cite{ilm92} found that the duration explained 80\% to 90\% of return variance of fixed rate government bonds during the 1980s. The nearly 100\% $R^2$-values for Model~1 during the past decade suggest that the trend has continued. It should be noted, however, that our results are not strictly comparable since our model explains the log-returns and it includes the time component.

\section{Inflation linked bonds}\label{sec:il}

The coupon and principal payments of most inflation linked bonds are tied to an underlying consumer price index so that the payment that will be received at time $t_n$ is $C_{t_n,n} = I_{t_n}C_{0,n}$, where $C_{0,n}$ is the {\em real payment} at time $t=0$ and $I_{t_n}$ is the value of the underlying consumer price index some time before $t_n$. The specification of the {\em indexation lag} is specified in the contract and depends e.g.\ on the lag between time for which the index is computed and the time when its value is reported. We refer the reader to \cite[Chapter~2]{ddm4} for a general overview of different cash flow structures on index linked bonds. The {\em real yield} of a portfolio of inflation linked bonds is defined by setting the outstanding payments according to the most recent value of the consumer price index $I_t$, i.e.
\[
C_{t,n} = I_tC_{0,n}
\]
in \eqref{y}. It follows that
\begin{align*}
\Delta C_{n} &= I_{s}C_{0,n} - I_{t}C_{0,n}\\
&= \left(\frac{I_{s}}{I_t}-1\right)C_{t,n}\\
&=\pi_{s}\Delta t C_{t,n},
\end{align*}
where $\pi_s$ is the {\em annualized rate of inflation} over the period $[t-\delta,s-\delta]$ where $\delta$ denotes the indexation lag. Assumption \eqref{k} thus holds with $K_s=\pi_s\Delta t$, so the first-order approximation \eqref{ret1} can be written as
\begin{equation}\label{eq:inflinked}
\Delta\ln P \approx (Y_s+\pi_s)\Delta t - D_t\Delta Y.
\end{equation}


\subsection{Empirical results}

In our empirical study, we consider the two model specifications
\begin{center}
\begin{tabular}{ll}
Model 1:	& $\Delta\ln P_s = c + Y_s\Delta t - D_t\Delta Y_s + \epsilon_s$, \\
Model 2:	& $\Delta\ln P_s = c + (Y_s+\pi_{s})\Delta t - D_t\Delta Y_s + \epsilon_s$,
\end{tabular}
\end{center}
the first one of which ignores the third term in the Taylor-approximation \eqref{ret0}. This allows us to evaluate the significance of the changes in the outstanding payments (the $K$-term in \eqref{ret1}) in explaining the portfolio returns. 

We use monthly observations of Barclays inflation linked government bond index data. The data consists of total return indices, yields and durations on portfolios of inflation linked government bonds for Canada, France, South Africa, Sweden, United Kingdom and United States\footnote{Further information is available online at https://ecommerce.barcap.com/indices/index.dxml}. The bonds' cash flows are linked to the evolution of country specific inflation indices (usually the general consumer price index). The monthly time series of appropriate inflation indices were obtained from Eurostat and the national statistical authorities' websites. Since the indexation lag is not specified in the data, we select the lag $\delta$ that gives the best fit to historical data. In all the studied countries, the data strongly supports a specific choice of the lag. The length of the time series for each country is given at the bottom of Table~\ref{tab:inflinked}. 

Table~\ref{tab:inflinked} gives the regression statistics for the two model specifications in the six markets. The inclusion of the $K$-term in Model 2 results in a substantial improvement over Model~1. Model~2 gives an extremely good fit to the total return data with $R^2$ values ranging from 97.6\% to 99.8\%. The lowest $R^2$ value for South Africa is mainly caused by one outlier observation caused by pricing distortions in the local conventional bond market in 2002 and the subsequent inversion of the breakeven inflation curve. 

\begin{figure}[!ht]
\centering
\subfigure[Canada]
{
		\epsfig{file=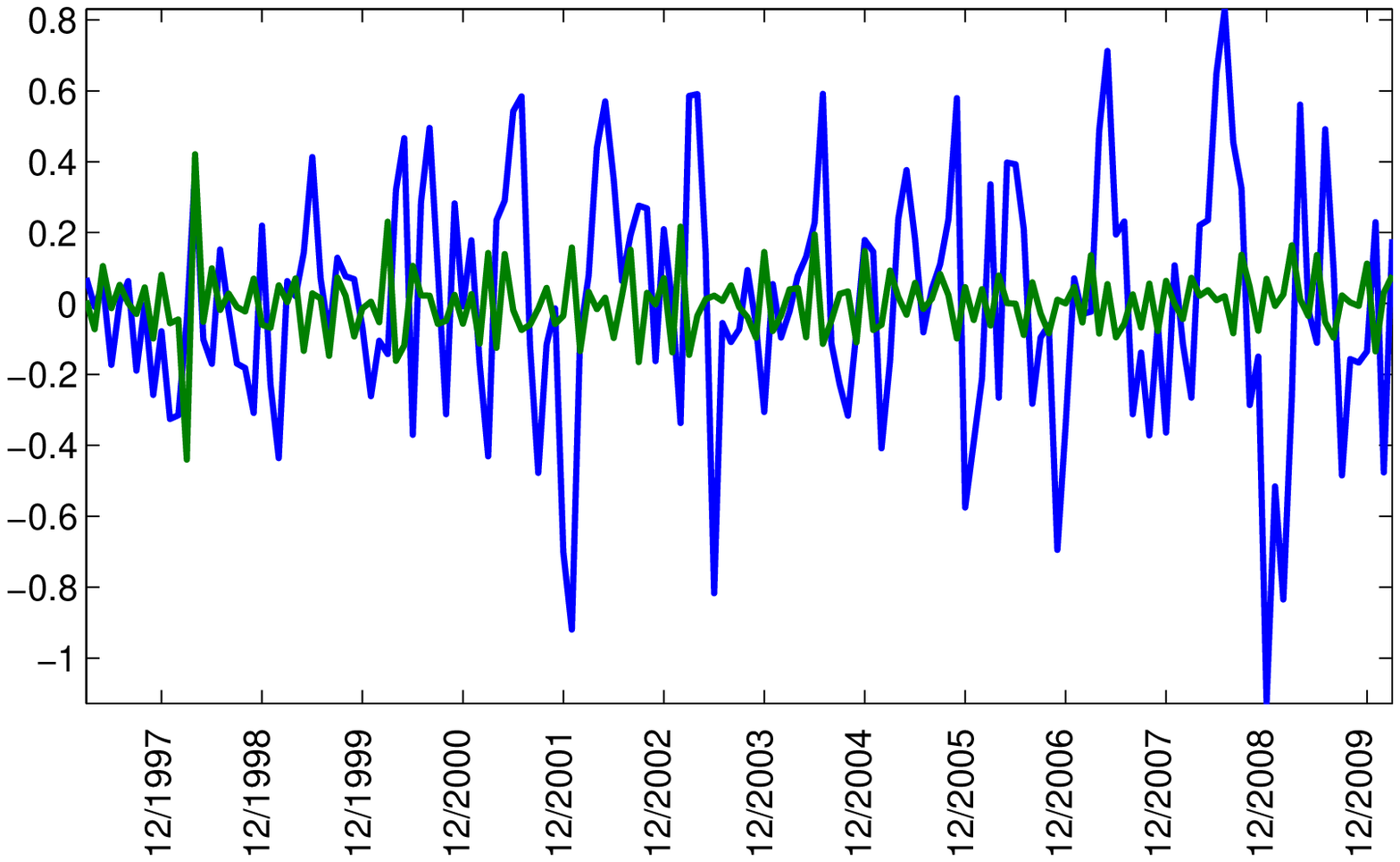,height=0.35\linewidth,width=0.47\linewidth,angle=0}
}
\subfigure[France]
{
		\epsfig{file=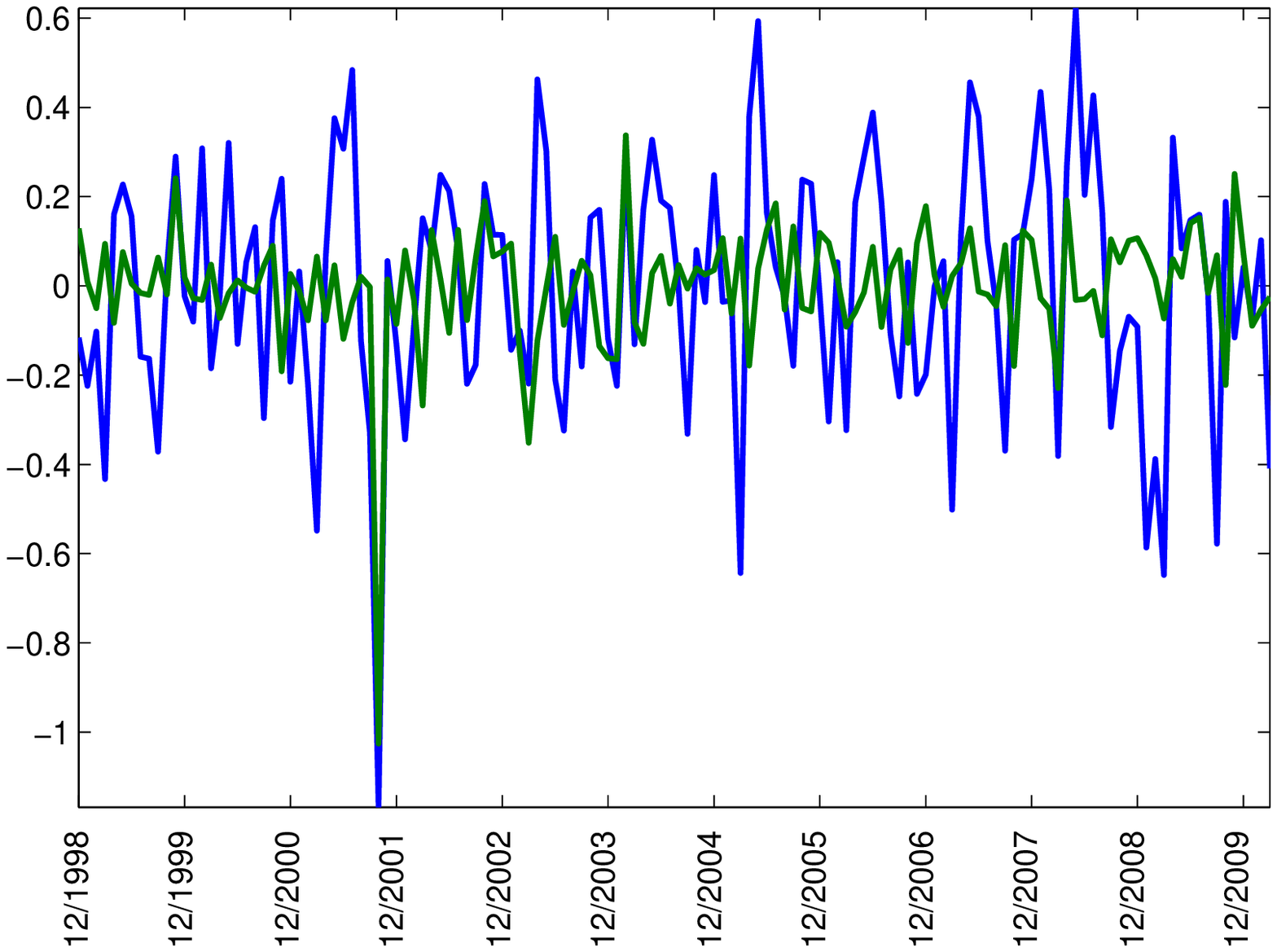,height=0.35\linewidth,width=0.47\linewidth,angle=0}
}
\subfigure[South Africa]
{
		\epsfig{file=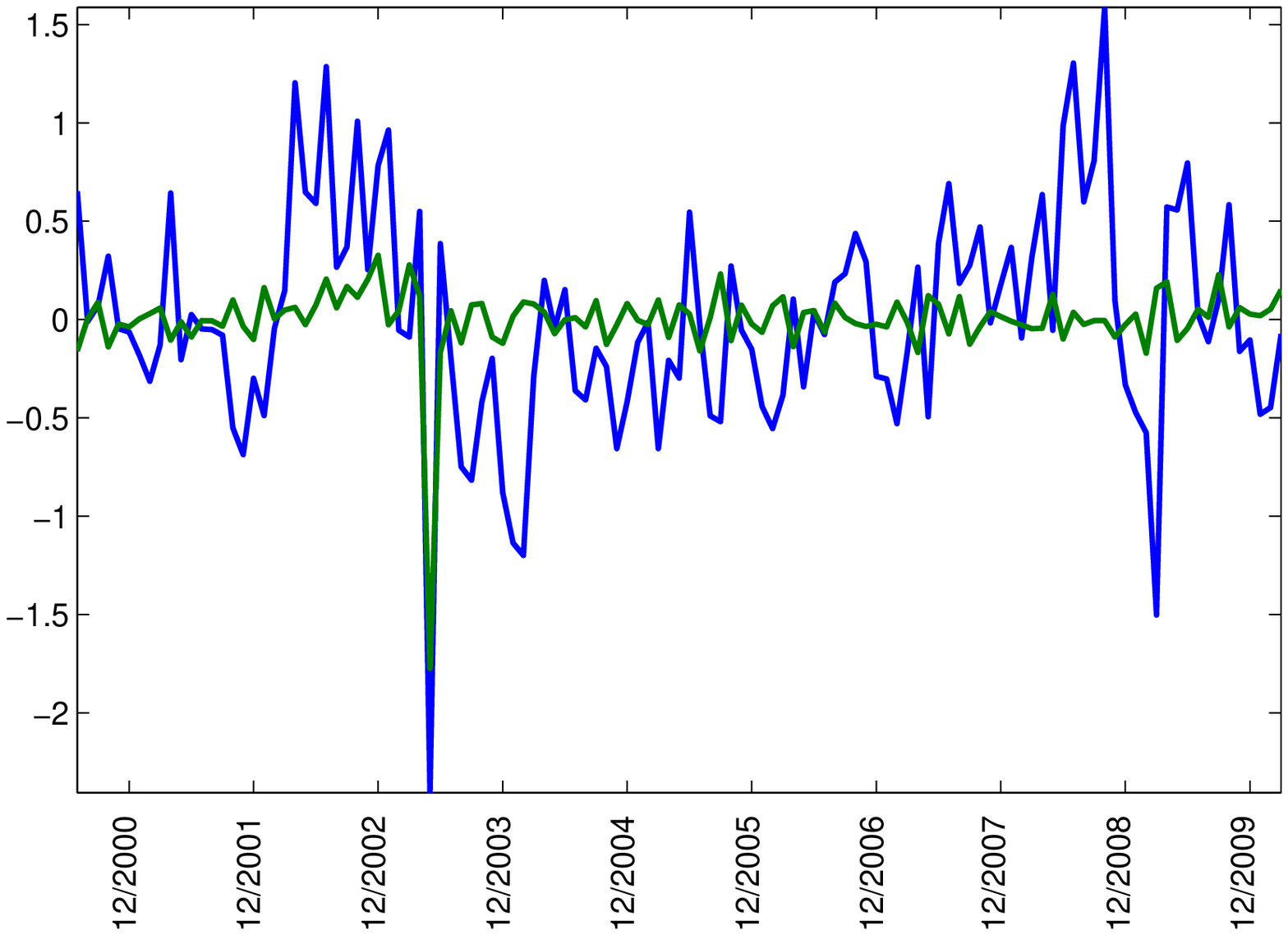,height=0.35\linewidth,width=0.47\linewidth,angle=0}
}
\subfigure[Sweden]
{
		\epsfig{file=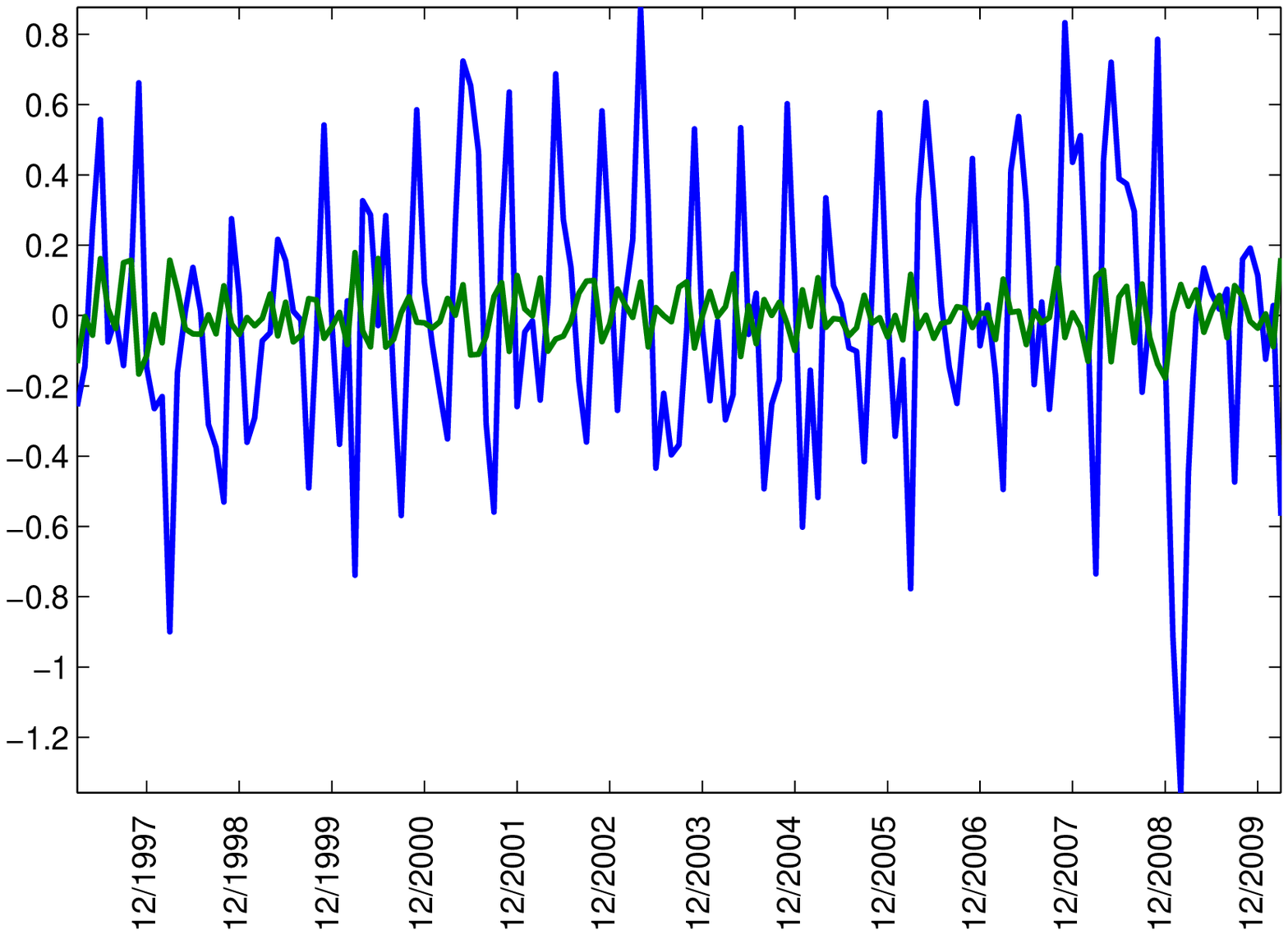,height=0.35\linewidth,width=0.47\linewidth,angle=0}
}
\subfigure[United Kingdom]
{
		\epsfig{file=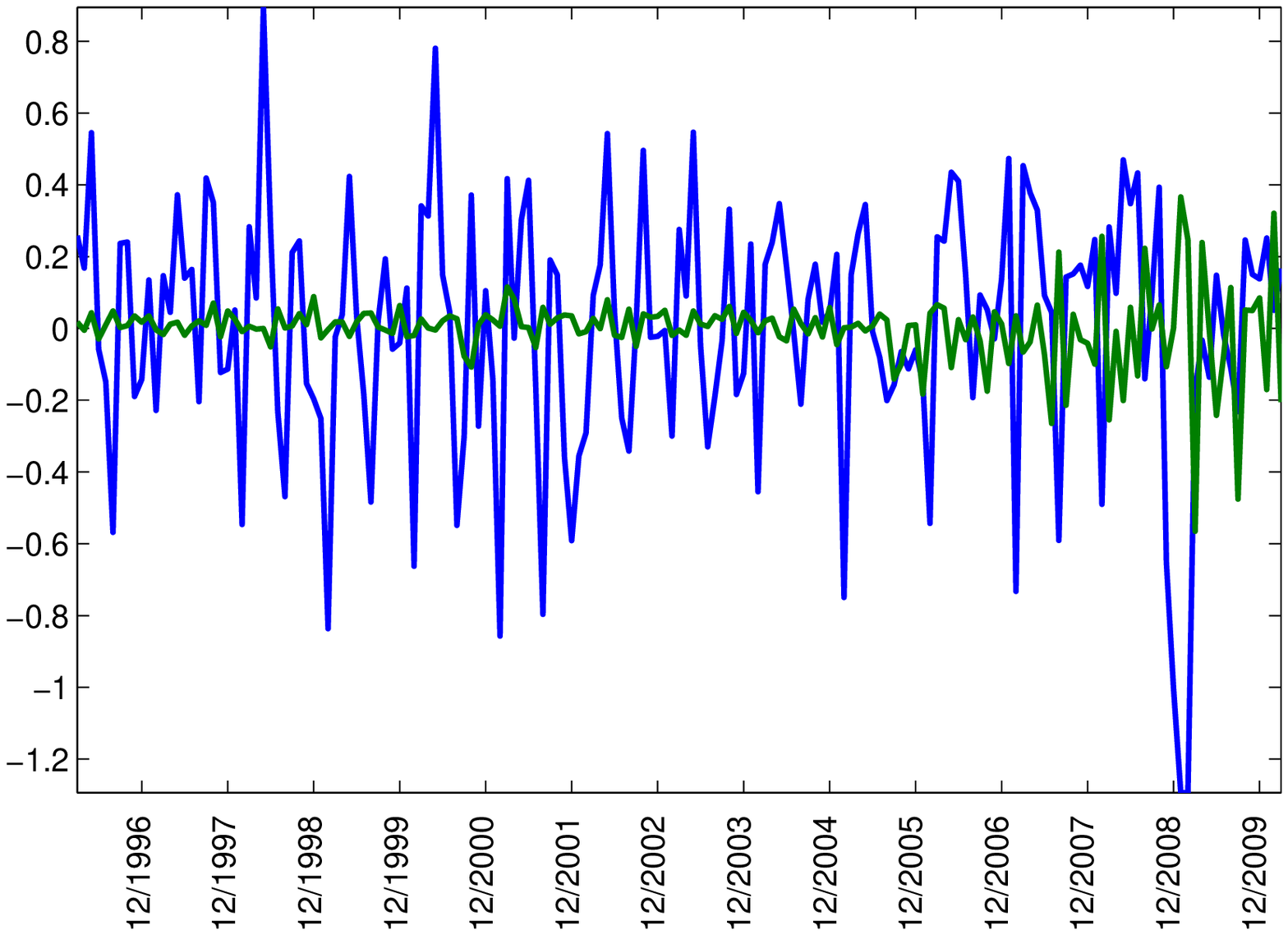,height=0.35\linewidth,width=0.47\linewidth,angle=0}
}
\subfigure[United States]
{
		\epsfig{file=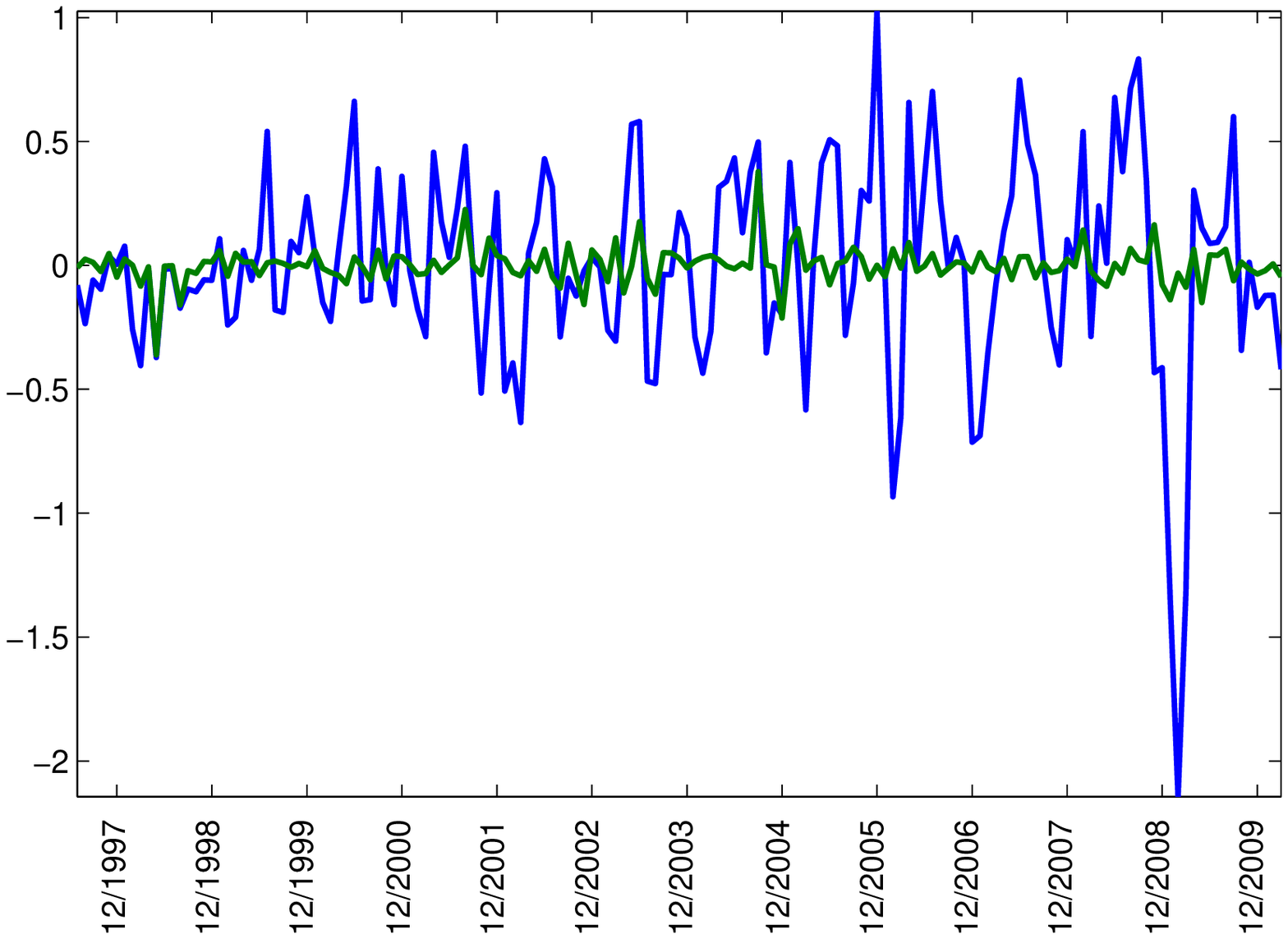,height=0.35\linewidth,width=0.47\linewidth,angle=0}
}
\caption{Residual returns (in percentages) of Model~1 (blue) and Model~2 (green) for inflation linked bonds.}
\label{fig:ILB_res} 
\end{figure}

The high values of Partial-$R^2$ statistics mean that the $K$-term is able to explain most of the residual variance of Model~1. Figure~\ref{fig:ILB_res} depicts the unexplained residual returns for the two models. The reduction in the residuals between Model 2 and Model 1 is caused solely by the incorporation of the inflation term. 

The fact that the estimated constant terms in Model~2 are generally insignificant is another indication that the model is well specified. The country specific ``optimal'' lag lengths $\delta$ are given on the last row of Table~\ref{tab:inflinked}.

\begin{table}[ht!]
\begin{center}
\caption{Regression statistics for inflation linked government bonds.}\label{tab:inflinked}
{\small
\begin{tabular}{r|cccccc}
\hline
&	CAN	&FRA	&SA &SWE	&UK	&US \\
\hline
\textbf{Model 1} &&&&&\\
100*c		&0.1642	&0.1328			&0.4935	&0.1114	&0.2073	&0.1966\\
t-stat 	&(6.201)	&(5.544)	&(9.367)	&(3.678)	&(7.611)	&(5.888)\\
$R^2$		&97.48\%	&95.46\%	&78.6\%	&91.29\%	&96.06\%	&94.41\%\\
\textbf{Model 2} &&&&&\\
100*c		&0.0046	&-0.0098	&-0.0069	&0.0087	&-0.0151	&-0.0004\\
t-stat 	&(0.628)&(-0.839)	&(-0.389)	&(1.465)	&(-1.918)	&(-0.066)\\
$R^2$		&99.80\%	&98.91\%	&97.58\%	&99.66\%	&99.67\%	&99.82\% \\
Partial-$R^2$	&92.22\%	&76.02\%	&88.69\%	&96.14\%	&91.43\%	&96.78\% \\
Lag	&2	&2	&3	&2	&1	&3\\
\hline
Data start	&1996-12	&1998-09		&2000-03 &1996-12 	&1996-01	&1997-03\\
Data end 	&2010-03	&2010-03		&2010-03	&2010-03 &2010-03	&2010-03\\
\hline
\multicolumn{7}{l}{ {\itshape{Notes.}} The table contains estimation results for the models:} \\
\multicolumn{7}{l}{ Model 1: $\Delta\ln P_s = c + Y_s\Delta t - D_t\Delta Y_s + \epsilon_s$,}   \\
\multicolumn{7}{l}{ Model 2: $\Delta\ln P_s = c + (Y_s+\pi_{s-k})\Delta t - D_t\Delta Y_s + \epsilon_s$,}\\
\multicolumn{7}{l}{ Numbers in parentheses are the $t$-statistics for the estimated coefficients.}
\end{tabular}}	
\end{center}
\end{table}

\section{Corporate bonds}\label{sec:corp}

It was found in \cite{imw94} that duration's explanatory power in explaining bond returns quickly decreases when moving from default-free bonds to corporate bonds. This section derives a simple two-factor model for the returns of well-diversified portfolios of non-callable corporate bonds. Our aim is to show that the effect of defaults can be approximated by \eqref{k} where $K_s$ can, in turn, be approximated by the yield spread between corporate and government bonds. We start by a mathematical justification based on familiar assumptions on the default process and on the diversification of the portfolio. The model will then be validated by an empirical analysis of Merril-Lynch data on corporate bond portfolios.

Because of defaults, the outstanding payments $C_{t,n}$ of a portfolio of corporate bonds may change during the holding period $[t,s]$. We will use the {\em multiple defaults} model described e.g.\ in \cite{sch98} and \cite[Chapter~6]{sch3}. The multiple defaults assumption can be justified by restructuring of defaulting issuers that often takes place in practice. Thus, we assume that each default event reduces the  outstanding payments $C_{t,n}$ to a fraction $\rho C_{t,n}$ of their pre-default values. The number $\rho$ is called the {\em recovery rate}, which in the context of portfolio modeling represents the reduction in the {\em portfolio's} outstanding payments due to an individual default. A crucial assumption in justifying \eqref{k}, is that each default event reduces the outstanding payments $C_{t,n}$ for all $n=1,\ldots,N$ by, approximately, the same recovery rate.

The time of occurrence and the recovery rate associated with the $i$th default will be denoted by $t_i$ and $\rho_i$, respectively. The sequence $(t_i,\rho_i)_{i=1}^\infty$ then determines the outstanding payments remaining after default events during a time interval $[t,s]$. Indeed, we have
\[
C_{s,n} = \prod_{i=1}^{I_s}\rho_iC_{t,n},
\]
where $I_s=\max\{i\,|\,t_i\in[t,s]\}$ is the number of default events during $[t,s]$. Both the default times $t_i$ and the recovery rates $\rho_i$ are random in general. In order to allow dependencies between them, we will model $(t_i,\rho_i)_{i=1}^\infty$ as a Cox process in $[0,\infty)\times[0,1]$ with intensity $\lambda_{t,\rho}$; see e.g.\ \cite{sch98} or \cite{gs8}. That is, we assume that, conditionally on $\lambda$, the number of default events with $(t_i,\rho_i)$ in a measurable subset $B$ of $[0,\infty)\times[0,1]$ is Poisson distributed with parameter
\[
\int_B\lambda_{t,\rho}dtd\rho
\]
and the number of events in disjoint subsets of $[0,\infty)\times[0,1]$ are independent. One could model defaults by a general {\em point process} (see e.g.\ \cite[Chapter~12]{kal2}), but for notational simplicity, we assume that  the process can be expressed in terms of an intensity. Note that the default times follow a Cox process with intensity $\lambda_t=\int_0^1\lambda_{t,\rho}d\rho$. The above model thus extends the one studied e.g.\ in \cite{lan98}; see \cite{sch98} for further discussion and references.


It follows (see the Appendix) that the amount of outstanding payments due at time $s\ge t$ has conditional expectation
\begin{equation}\label{ec}
E[C_{s,n}\,|\,\lambda] = C_{t,n}\exp\left(-\int_t^sl_udu\right),
\end{equation}
where 
\[
l_u=\int_0^1(1-\rho)\lambda_{u,\rho}d\rho
\]
is the {\em mean loss rate}. Note that we can write $l_t=\lambda_t(1-\rho_t)$, where $\lambda_t=\int_0^1\lambda_{t,\rho}d\rho$ and $\rho_t=\int_0^1\rho\lambda_{t,\rho}d\rho/\lambda_t$ for $\lambda_t>0$ and $\rho_t=0$ otherwise. This corresponds to the defaultable term structure models proposed e.g.\ in \cite{ds99}.

Arguing like in \cite{jly5} (who studied defaultable zero coupon bonds with zero recovery), one can show that the default event risk can be diversified away in the sense that the outstanding payments $C_{s,n}$ converge to $E[C_{s,n}\,|\,\lambda]$ when the number of issuers in the portfolio is increased; see the Appendix. This suggests that for a well-diversified portfolio, the effect of default events on outstanding payments can be approximated by
\begin{equation}\label{ec0}
C_{s,n} \approx C_{t,n}\exp\left(-\int_t^sl_udu\right)
\end{equation}
and thus that \eqref{k} holds with 
\[
K_s = \exp\left(-\int_t^sl_udu\right) - 1.
\]
In order to estimate $K_s$, we will assume that there exists a {\em risk neutral measure} $Q$ under which the market prices of traded securities are equal to the expectations of their discounted cash flows; see e.g.\ \cite[Chapter~5]{ds3}. The default-free short rate will be denoted by $r$. The amount of cash that will be received at time $t_n$ is equal to $C_{t_n,n}$, the outstanding payments at time $t_n$ payable at time $t_n$. Applying \eqref{ec0} with $s=t_n$, we get
\begin{align*}
P_t &= E^Q\sum_{n=1}^N\exp\left(-\int_t^{t_n}r_udu\right) C_{t_n,n}\\
&\approx E^Q\sum_{n=1}^N\exp\left(-\int_t^{t_n}r_udu\right)\exp\left(-\int_t^{t_n}l_udu\right)C_{t,n}\\
&= E^Q\sum_{n=1}^N\exp\left(-\int_t^{t_n}r^c_udu\right)C_{t,n},
\end{align*}
where $r^c_t=r_t+l_t$ is the {\em defaultable short rate}. This is analogous to the reduced form pricing formulas obtained e.g.\ in \cite{sch98,ds99}. The instantaneous loss rate can thus be expressed as the {\em short spread} $l_t=r^c_t-r_t$. 

\begin{figure}[!ht]
\centering
\subfigure[Maturity 1-3 years]
{
		\epsfig{file=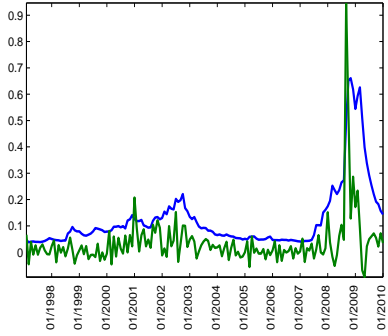,height=0.4\linewidth,width=0.47\linewidth,angle=0}
}
\subfigure[Maturity 3-5 years]
{
		\epsfig{file=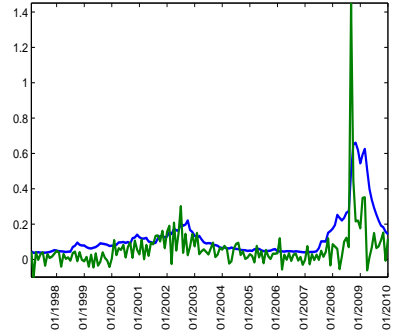,height=0.4\linewidth,width=0.47\linewidth,angle=0}
}
\subfigure[Maturity 5-7 years]
{
		\epsfig{file=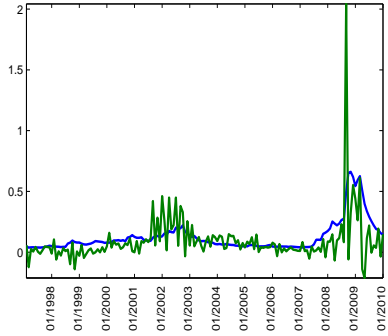,height=0.4\linewidth,width=0.47\linewidth,angle=0}
}
\subfigure[Maturity 7-10 years]
{
		\epsfig{file=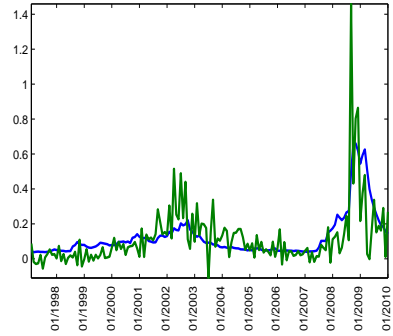,height=0.4\linewidth,width=0.47\linewidth,angle=0}
}
\caption{Historical evolution (in percentages) of $S_t\Delta t$ (blue) and residual returns of Model 1 representing default losses (green) during 1996/2--2010/1 for the US indices.}
\label{fig:US_S_L} 
\end{figure}

\begin{figure}[!ht]
\centering
\subfigure[Maturity 1-3 years]
{
		\epsfig{file=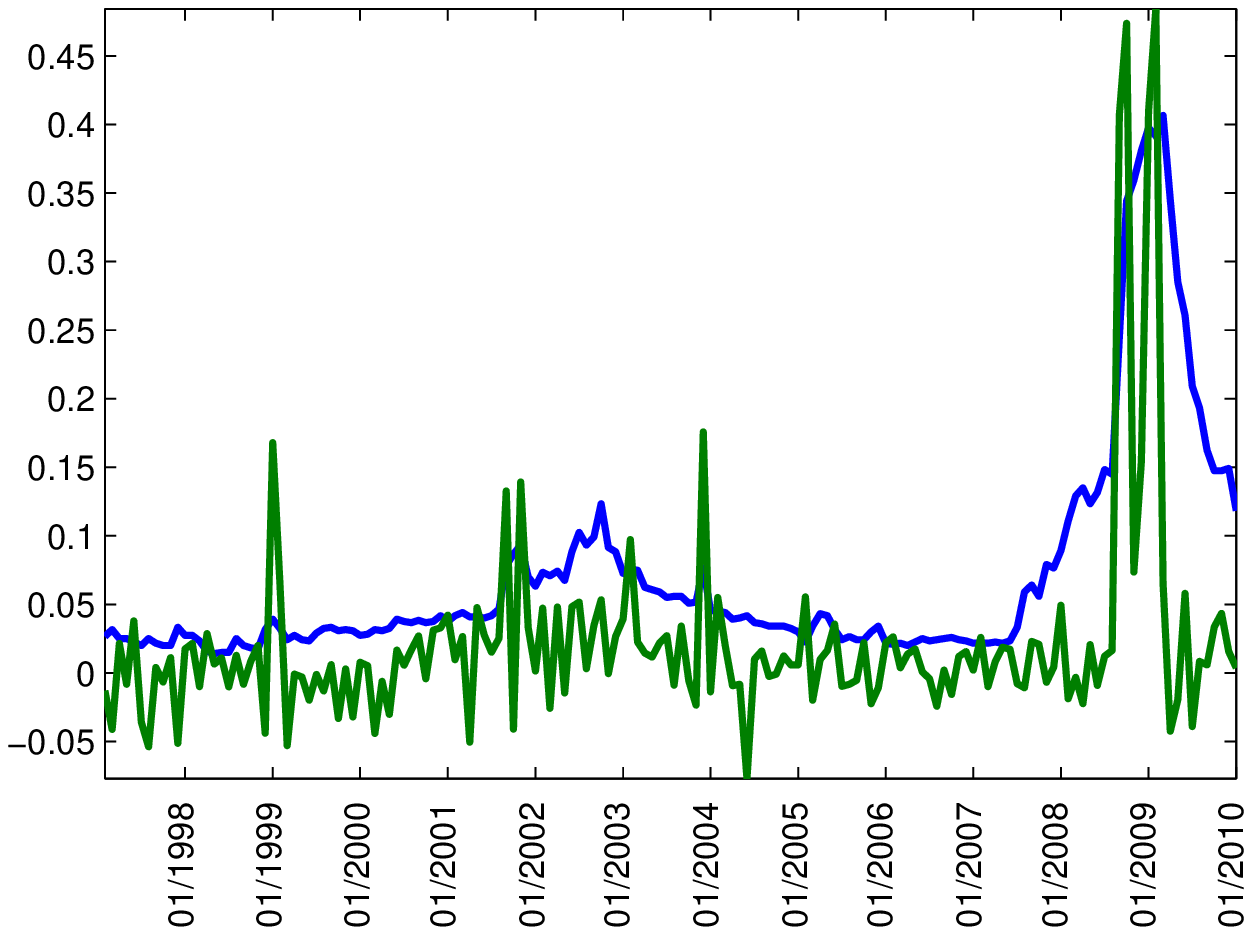,height=0.4\linewidth,width=0.47\linewidth,angle=0}
}
\subfigure[Maturity 3-5 years]
{
		\epsfig{file=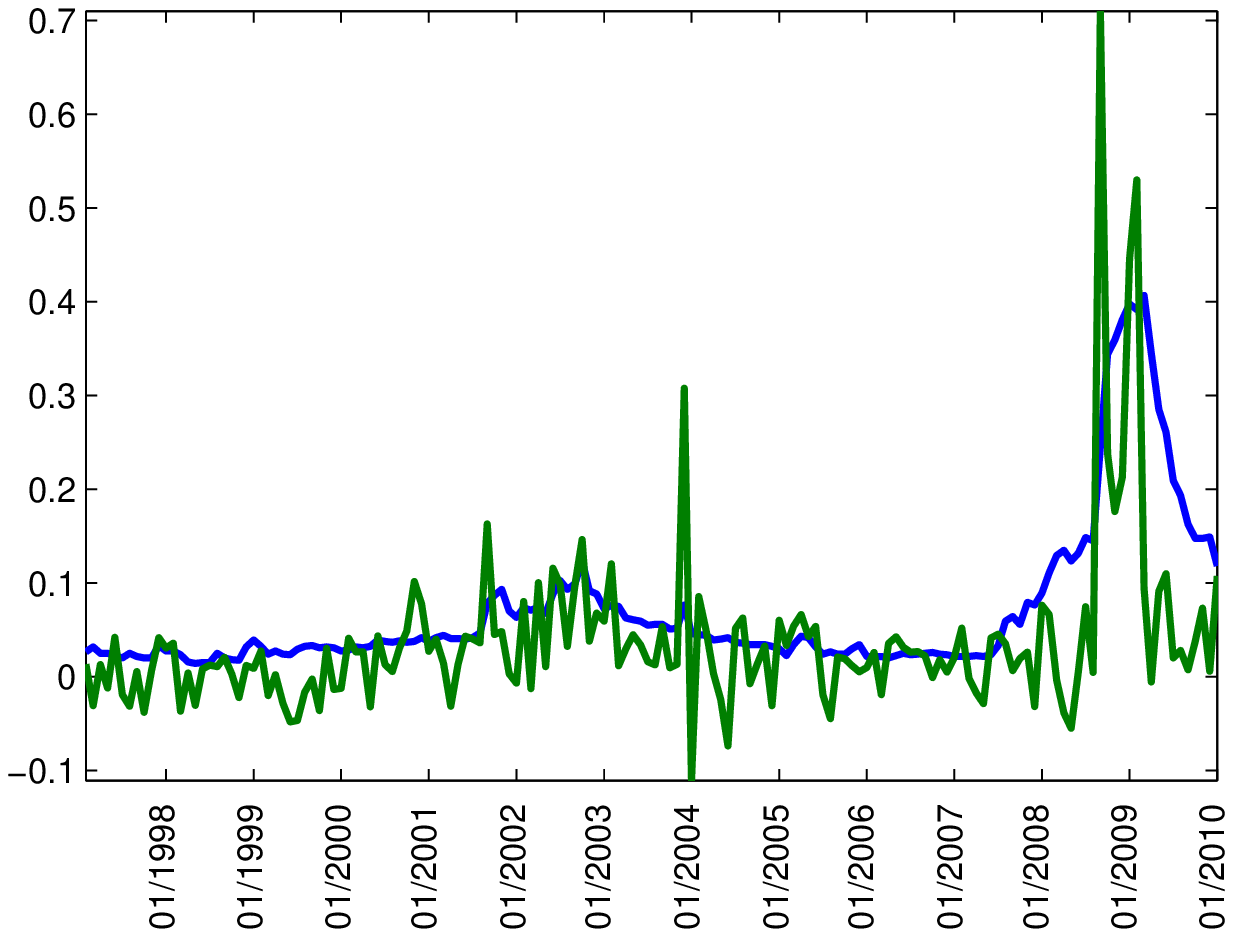,height=0.4\linewidth,width=0.47\linewidth,angle=0}
}
\subfigure[Maturity 5-7 years]
{
		\epsfig{file=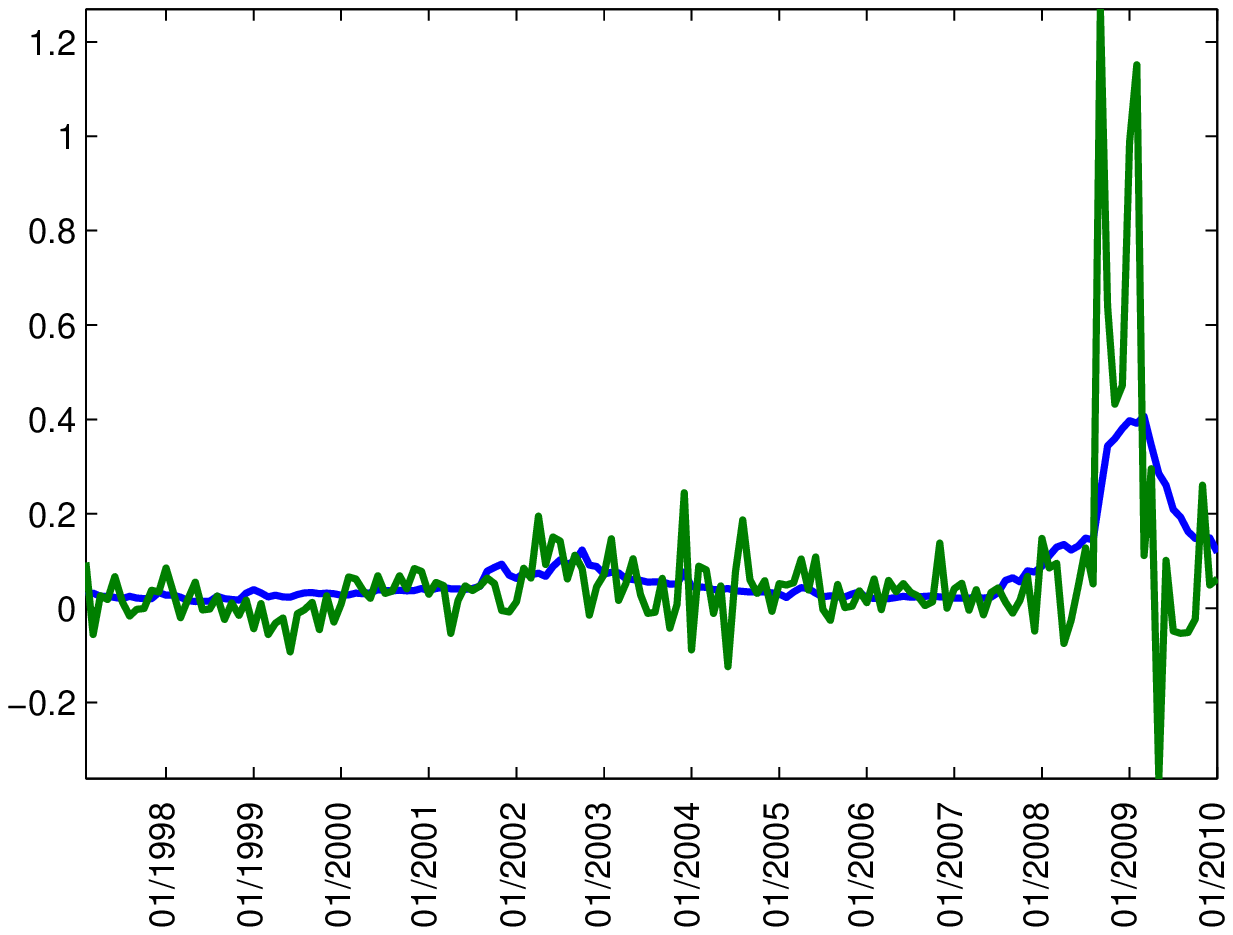,height=0.4\linewidth,width=0.47\linewidth,angle=0}
}
\subfigure[Maturity 7-10 years]
{
		\epsfig{file=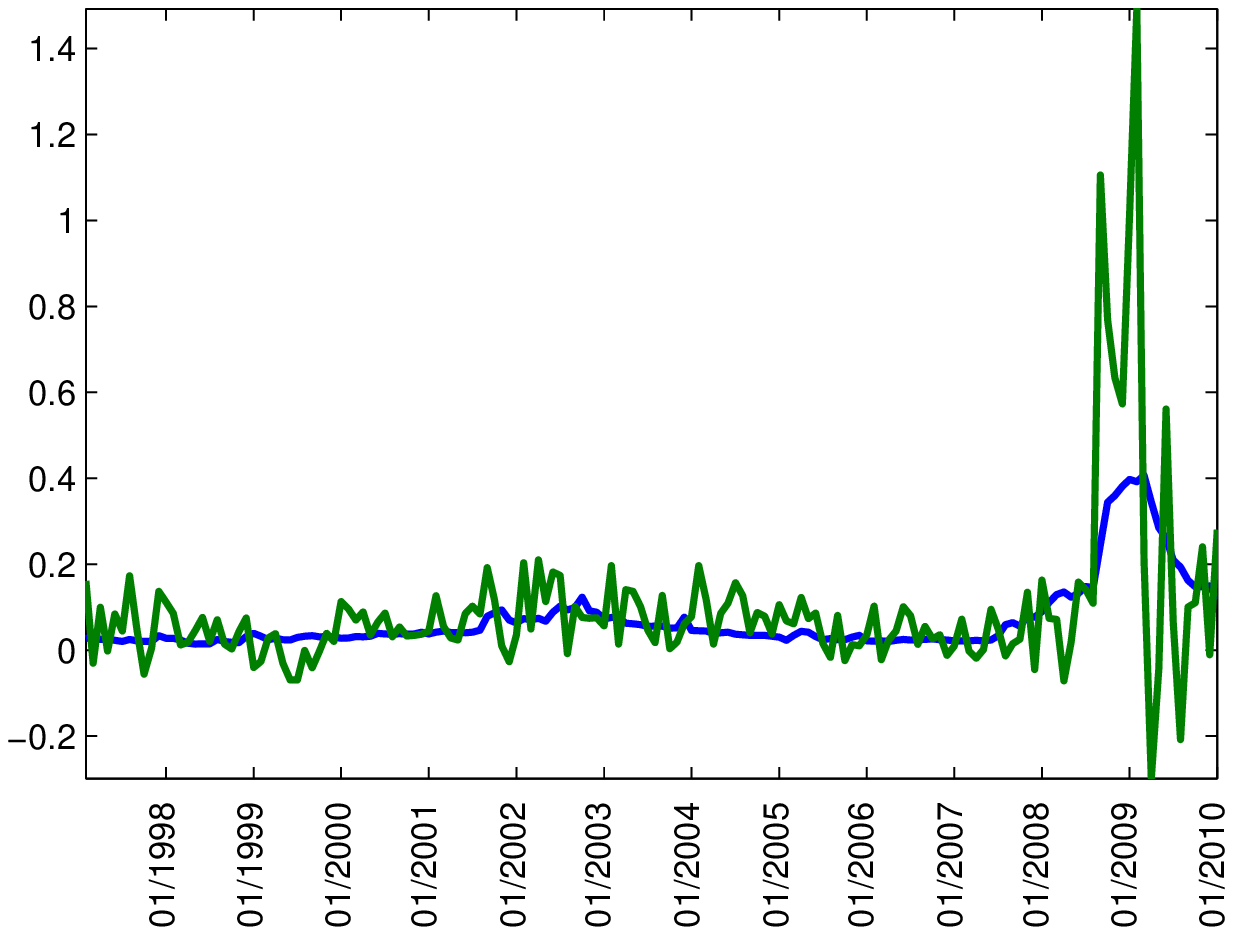,height=0.4\linewidth,width=0.47\linewidth,angle=0}
}
\caption{Historical evolution of $S_t\Delta t$ (blue) and residual returns of Model 1 representing default losses (green) during 1996/2--2010/1 for the European indices.}
\label{fig:EU_S_L} 
\end{figure}

Unfortunately, neither $r^c_t$ nor $r_t$ is observable in practice. We will approximate them by yields on short maturity corporate and government bond portfolios, respectively. This suggests the approximation
\begin{equation}\label{kexp}
K_s\approx \exp(- S_s\Delta t)-1,
\end{equation}
where $S_s$ is the short spread at time $s$. When $S_s\Delta$ is small, we can simplify this to
\begin{equation}\label{ks}
K_s\approx -S_s\Delta t.
\end{equation}
Figures~\ref{fig:US_S_L} and \ref{fig:EU_S_L} display historical monthly values of $K_s$ and $S_s\Delta t$ in percentages for European and US corporate bond markets during 1996/2--2010/1. The values of $K_s$ were obtained by solving $K_s$ from \eqref{ret1} for observed values of log-returns and yields on Merrill-Lynch investment grade corporate bond indices. In the figures, the spread $S_s$ is the difference between yields on portfolios of corporate and government bonds of 1-3 years to maturity. The values of $S_s\Delta t$ have remained below $0.7\%$ which means that the relative error in using \eqref{ks} instead of \eqref{kexp} would have been less than $0.0025\%$ points. 

Using \eqref{ks} in \eqref{ret1} suggests the approximation
\begin{equation}\label{retc}
\Delta\ln P\approx (Y_s-\alpha S_s)\Delta t - D_t\Delta Y
\end{equation}
for log-returns on corporate bond portfolios. We allow the constant $\alpha$ to deviate from one since even with the shortest maturity available (1-3 years in our numerical examples below), the spread $S_s$ is likely to overestimate the mean loss rate $l_s$. Indeed, the yield on corporate bonds contains a premium for unexpected variations in the future loss rate. In addition to this {\em systematic default risk} there may be a premium for {\em default event risk} which remains when the diversification argument used to derive \eqref{ec0} is not exactly satisfied; see the Appendix.

\subsection{Empirical results}

We will again consider two different model specifications
\begin{center}
\begin{tabular}{ll}
Model 1:	& $\Delta\ln P_s = c + Y_s\Delta t - D_t\Delta Y_s + \epsilon_s$, \\
Model 2:	& $\Delta\ln P_s = c + (Y_s-\alpha S_s)\Delta t - D_t\Delta Y_s + \epsilon_s$, 
\end{tabular}
\end{center}
which differ by the term $\alpha S_s$, where $S_s$ is the yield spread on short maturity bond indices. The parameter $\alpha$ will be estimated from the data. We fit the models to monthly observations of Merrill-Lynch investment grade corporate bond indicex data. Our dataset covers the total return indices, yields and durations of market capitalization weighted investment grade corporate bond portfolios of maturities 1--3, 3--5, 5--7 and 7--10 years from the US and European markets. As a proxy for the instantaneous loss rate we will use the yield spread of 1--3 year short maturity bonds.

\begin{table}[!ht]
\begin{center}
{\small
\caption{Return regressions for the EU corporate bond portfolios.}\label{tab:EU_est_res_corp}
\begin{tabular}{c|cccc|cccc}
\hline
 & \multicolumn{4}{c|}{Model 1} & \multicolumn{4}{c}{Model 2}  \\
 \hline
	&1-3 Y	&3-5 Y	&5-7 Y	&7-10 Y	&1-3 Y	&3-5 Y	&5-7 Y	&7-10 Y \\
\hline
$100*c$	&-0.022	&-0.038	&-0.064	&-0.094	&0.013	&0.007	&0.028	&0.012\\
				&(-3.431)	&(-5.014)	&(-4.357)	&(-5.754)	&(1.811)	&(0.818)	&(1.753)	&(0.675)\\
$\alpha$	&	&	&	&	&0.509	&0.633	&1.312	&1.502\\
					&	&	&	&	&(7.424)	&(8.246)	&(8.921)	&(9.426)\\
$R^2$	&96.38\%	&98.63\%	&97.5\%	&98.21\%	&97.33\%	&99.05\%	&98.35\%	&98.87\%\\
Partial-$R^2$	&	&	&	&	&26.35\%	&30.63\%	&34.07\%	&36.58\% \\
\hline
Data start	&1997-1	&1997-1	&1997-1	&1997-1	&1997-1	&1997-1	&1997-1	&1997-1\\
Data end	&2010-1	&2010-1	&2010-1	&2010-1	&2010-1	&2010-1	&2010-1	&2010-1\\
\hline
\multicolumn{9}{l}{ {\itshape{Notes.}} The table contains estimation results for two different model specifications:} \\
\multicolumn{9}{l}{ Model 1: $\Delta\ln P_s = c + Y_s\Delta t - D_t\Delta Y_s + \epsilon_s$,}   \\
\multicolumn{9}{l}{ Model 2: $\Delta\ln P = (Y_s-\alpha S_s)\Delta t - D_t\Delta Y_s + \epsilon_s$,}\\
\multicolumn{9}{l}{ Numbers in parentheses are the $t$-statistics for the estimated coefficients.}
\end{tabular}
}
\end{center}
\end{table}

\begin{table}[!ht]
\begin{center}
\caption{Return regressions for the US corporate bond portfolios.}\label{tab:US_est_res_corp}
{\small
\begin{tabular}{c|cccc|cccc}
\hline
 & \multicolumn{4}{c|}{Model 1} & \multicolumn{4}{c}{Model 2} \\
 \hline
	&1-3 Y	&3-5 Y	&5-7 Y	&7-10 Y	&1-3 Y	&3-5 Y	&5-7 Y	&7-10 Y	\\
\hline
$100*c$	&-0.033	&-0.061	&-0.09	&-0.116	&0.02	&0.011	&-0.001	&-0.001\\
				&(-3.506)	&(-5.265)	&(-5.424)	&(-8.145)	&(1.754)	&(0.747)	&(-0.035)	&(-0.098)\\
$\alpha$	&	&	&	&	&0.423	&0.571	&0.712	&0.915\\
					&	&	&	&	&(6.462)	&(7.169)	&(6.005)	&(10.643)\\
$R^2$	&97.81\%	&98.51\%	&98.35\%	&99.19\%	&98.27\%	&98.88\%	&98.66\%	&99.53\%\\
Partial-$R^2$	&	&	&	&	&21.33\%	&25.02\%	&18.97\%	&42.38\% \\
\hline
Data start	&1997-1	&1997-1	&1997-1	&1997-1	&1997-1	&1997-1	&1997-1	&1997-1\\
Data end	&2010-1	&2010-1	&2010-1	&2010-1	&2010-1	&2010-1	&2010-1	&2010-1\\
\hline
\multicolumn{9}{l}{ {\itshape{Notes.}} The table contains estimation results for two different model specifications:} \\
\multicolumn{9}{l}{ Model 1: $\Delta\ln P_s = c + Y_s\Delta t - D_t\Delta Y_s + \epsilon_s$,}   \\
\multicolumn{9}{l}{ Model 2: $\Delta\ln P = (Y_s-\alpha S_s)\Delta t - D_t\Delta Y_s + \epsilon_s$,}\\
\multicolumn{9}{l}{ Numbers in parentheses are the $t$-statistics for the estimated coefficients.}
\end{tabular}}	
\end{center}
\end{table}

Table~\ref{tab:EU_est_res_corp} displays the estimation results for the European and Table~\ref{tab:US_est_res_corp} for the US data under the two model specifications. Model 1, which ignores the effects of default losses on the outstanding payments, already explains the majority of the return variations with $R^2$ values ranging between 96.4\% and 99.2\%. The addition of the the spread term in Model 2 inmproves the model fit in all maturities and markets. The spread term is able to explain roughly 20--40\% of the residual return variation of Model~1. In addition to improving the models' fit the addition of the spread terms also dilutes the significance of the estimated constant terms, which may be interpreted as a sign of an improved model specification. These quantitative results are well supported by Figures \ref{fig:US_S_L} and \ref{fig:EU_S_L} that display the evolution of unexplained residual returns of Model~1 together with the short maturity yield spread in the US and European markets, respectively.

\begin{figure}[!ht]
\centering
\subfigure[Maturity 1-3 years]
{
		\epsfig{file=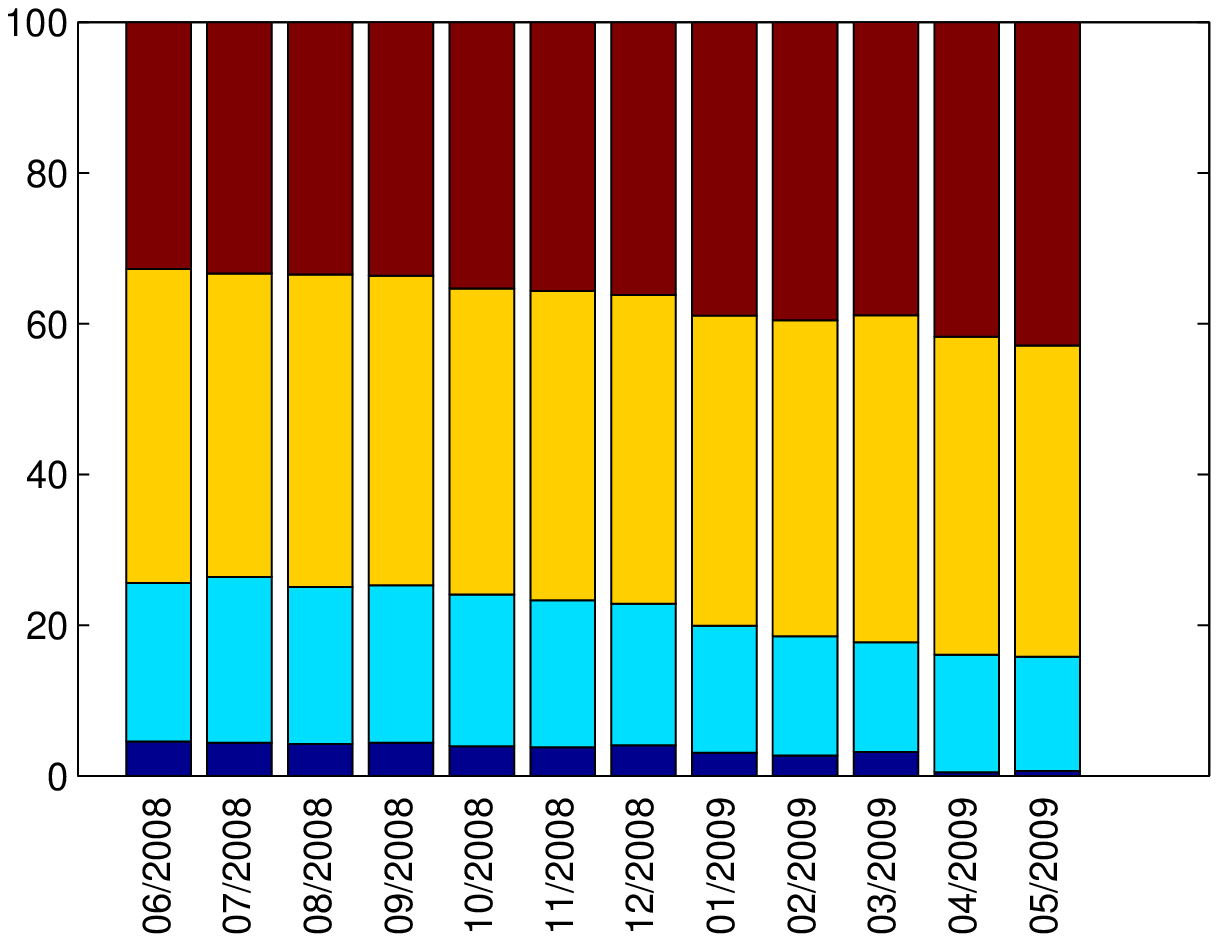,height=0.4\linewidth,width=0.47\linewidth,angle=0}
}
\subfigure[Maturity 3-5 years]
{
		\epsfig{file=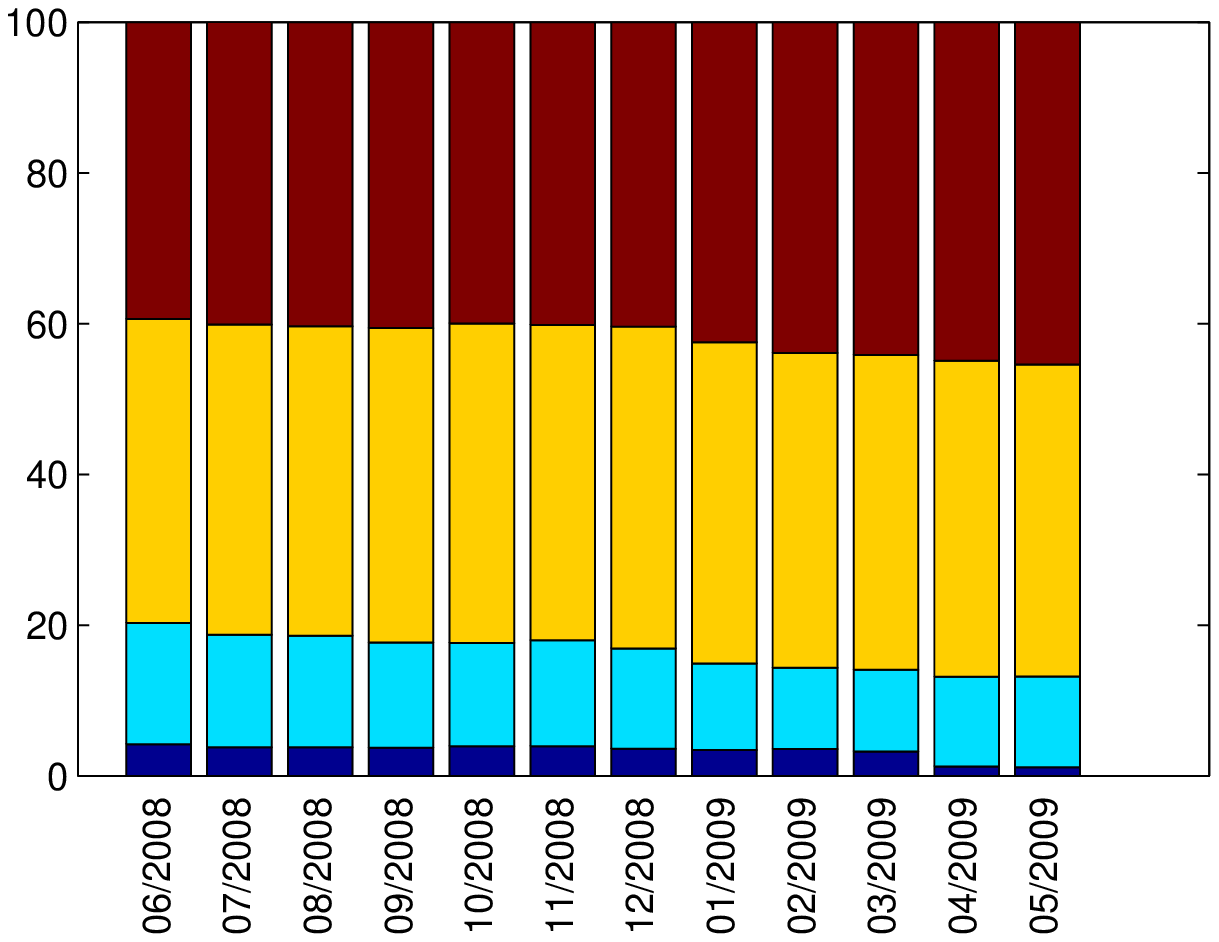,height=0.4\linewidth,width=0.47\linewidth,angle=0}
}
\subfigure[Maturity 5-7 years]
{
		\epsfig{file=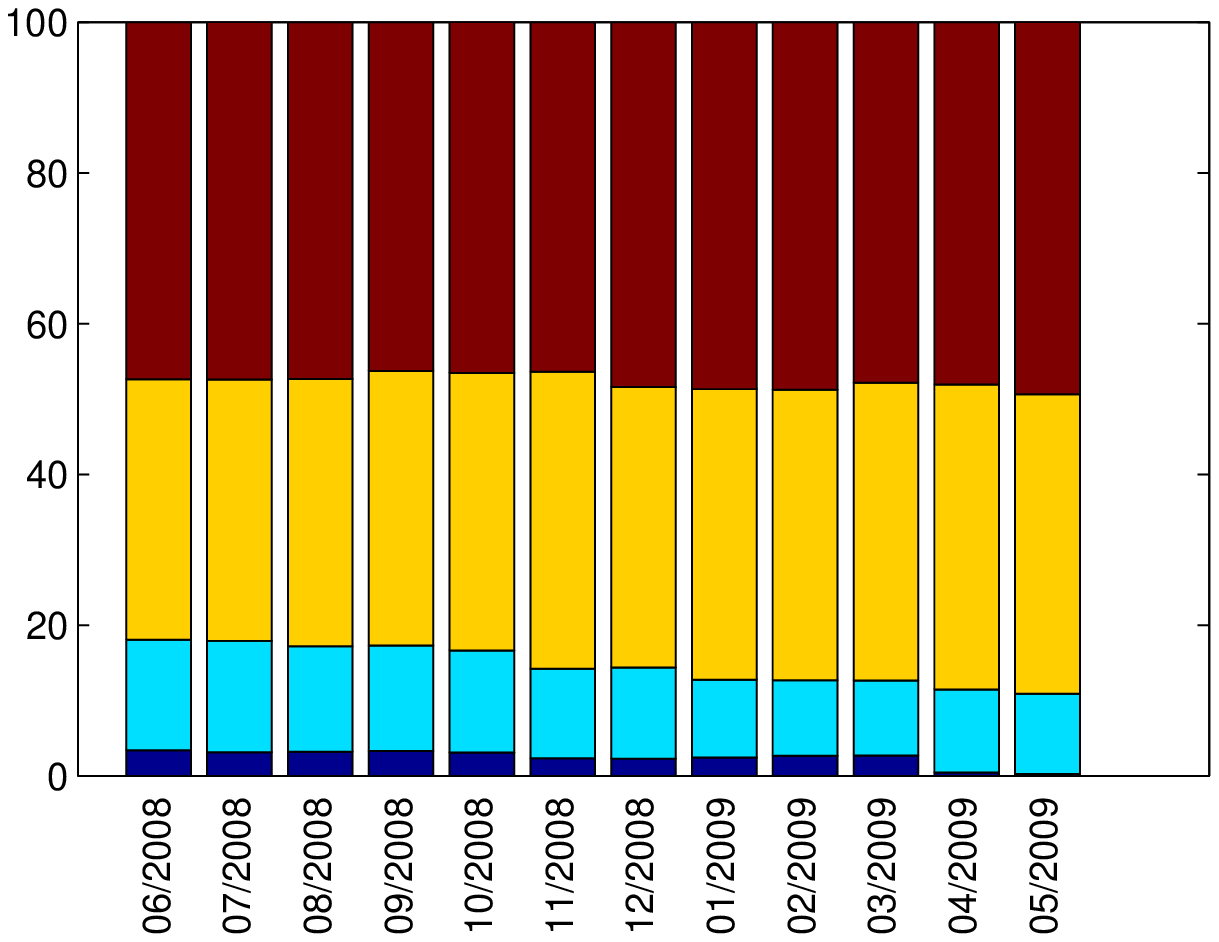,height=0.4\linewidth,width=0.47\linewidth,angle=0}
}
\subfigure[Maturity 7-10 years]
{
		\epsfig{file=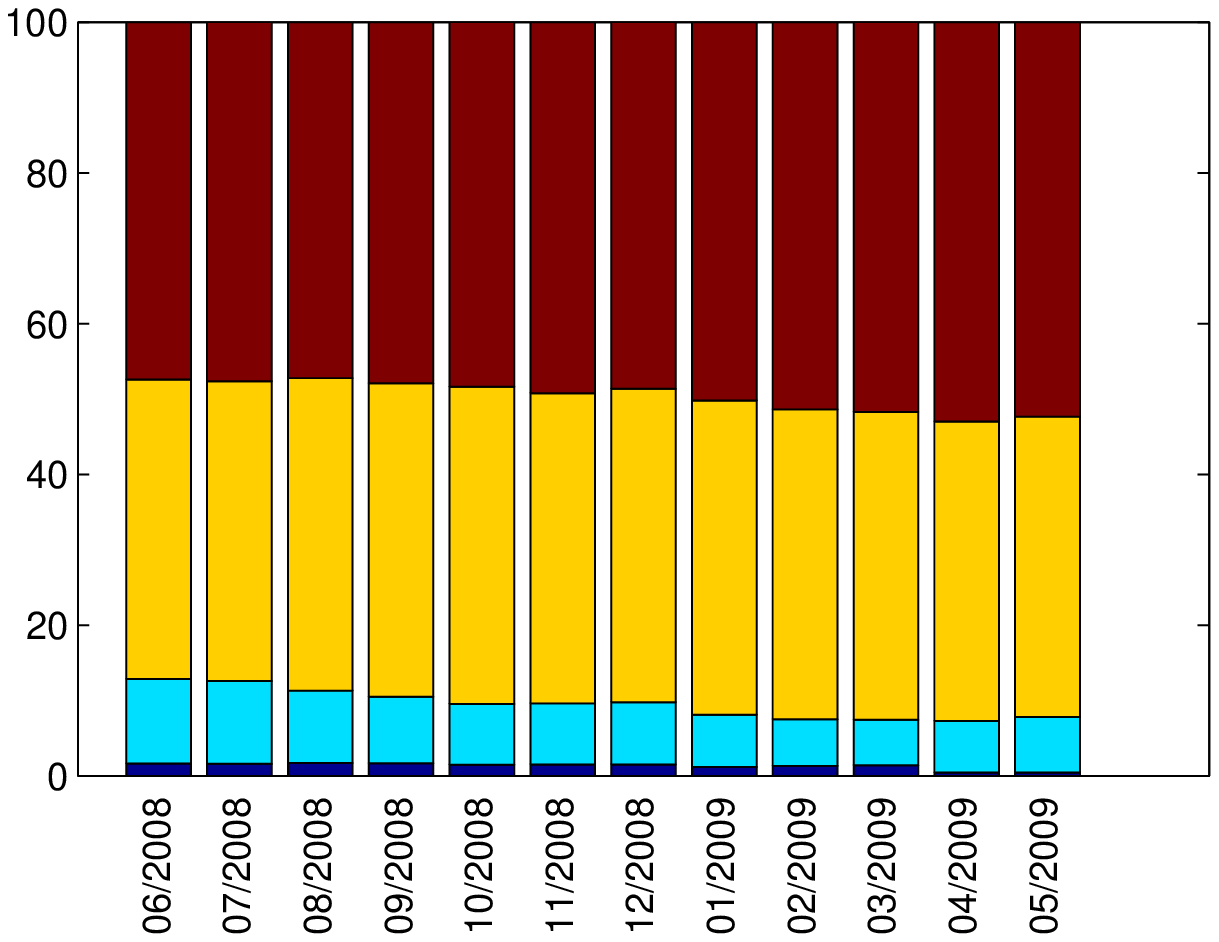,height=0.4\linewidth,width=0.47\linewidth,angle=0}
}
\caption{The composition of US corporate bond indices by rating classes AAA (dark blue), AA (light blue), A (yellow) and BBB (red) during 6/2009-5/2010. }
\label{fig:US_rating_distr} 
\end{figure}

\begin{figure}[!ht]
\centering
\subfigure[Maturity 1-3 years]
{
		\epsfig{file=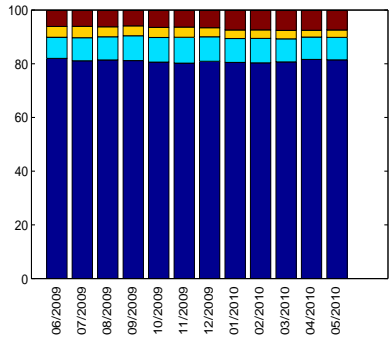,height=0.4\linewidth,width=0.47\linewidth,angle=0}
}
\subfigure[Maturity 3-5 years]
{
		\epsfig{file=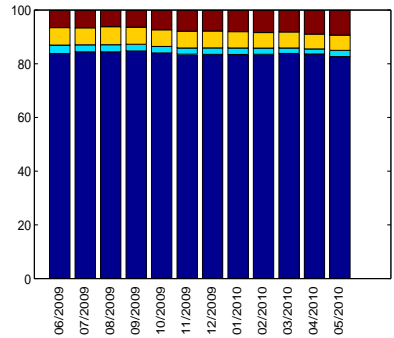,height=0.4\linewidth,width=0.47\linewidth,angle=0}
}
\subfigure[Maturity 5-7 years]
{
		\epsfig{file=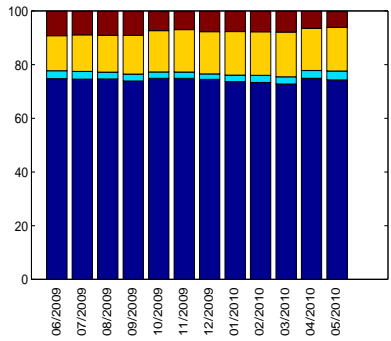,height=0.4\linewidth,width=0.47\linewidth,angle=0}
}
\subfigure[Maturity 7-10 years]
{
		\epsfig{file=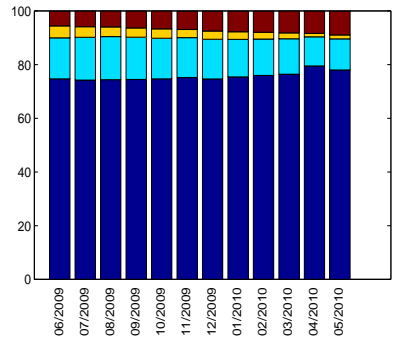,height=0.4\linewidth,width=0.47\linewidth,angle=0}
}
\caption{The distribution of US corporate bond indices among senior debt (dark blue), subordinated debt (light blue), capital instruments (yellow) and securitized debt (red) during 6/2009-5/2010. }
\label{fig:US_distr} 
\end{figure}

The estimates of $\alpha$ as well as the Partial-$R^2$ values tend to increase as a function of the index maturity. This can be explained, to a large extent, by the differences in the rating composition and seniority of the bonds included in the indices. Figure \ref{fig:US_rating_distr} displays the evolution of the rating distributions of the US corporate bond indices by maturity bucket during the worst phases of the financial crises, when the default losses reached exceptionally high levels. The rating quality deteriorates with maturity which explains the increased default losses at the longer maturities. The composition of the US indices by instrument type are displayed in Figure \ref{fig:US_distr}. In line with the deterioration of the rating quality, the seniority of the bonds contained in the indices also declines as the maturity of the indices increases. The largest default losses were observed in the 5--7 year index which had the largest share of capital related debt instruments (under the Basel II regulation) and securitized exposures, both of which exhibited large losses during the financial crisis. On the other hand, the shorter end of the maturity spectrum contained the highest share of senior debt and thus exhibited the lowest default losses. Similar conclusions regarding the deterioration of the rating quality of the indices as a function of the maturity hold for the European indices as well but we omit the details.  

The three terms in Model~2 explain majority of the return variation in the considered eight portfolios but, especially at the longer maturities, there remains residual spikes in the aftermath of the Lehman Brothers' collapse in September 2008. Some idiosyncratic default event risk thus seems to remain, even though the indices contain bonds from hundreds of issuers. The systematic default risk on the other hand seems to be captured to a large extent by the spread term.

\section*{Appendix}

We prove \eqref{ec} by using the notion of {\em Laplace functional} of a random measure; see e.g.\ \cite[Chapter~12]{kal2}. To this end, we write
\begin{align*}
C_{s,n} &= \prod_{i=1}^{I_s}\rho_iC_{t,n} \\
&= \exp\left(\sum_{i=1}^{I_s}\ln\rho_i\right)C_{t,n} \\
&= \exp\left(-\int_t^\infty\int_0^1 fd\mu(u,\rho)\right)C_{t,n},
\end{align*}
where $f(u,\rho)=-\chi_{[t,s]}(u)\ln\rho$ and $\mu=\sum_i\delta_{t_i,\rho_i}$. Here $\chi_{[t,s]}$ denotes the characteristic function of the interval $[s,t]$ and $\delta_{t_i,\rho_i}$ is the Dirac measure that assigns unit mass to the point $(t_i,\rho_i)$. Conditionally on $\lambda$, the process $(t_i,\rho_i)_{i=1}^\infty$ is Poisson (since it is a Cox process), so by \cite[Lemma~12.2(i)]{kal2},
\begin{align*}
E[C_{s,n}\,|\,\lambda] &= \exp\left(-\int_t^\infty\int_0^1(1-e^{-f(u,\rho)})\lambda_{u,\rho}dud\rho\right)C_{t,n}\\
&= \exp\left(-\int_t^s\int_0^1(1-\rho)\lambda_{u,\rho}dud\rho\right)C_{t,n}\\
&= \exp\left(-\int_t^sl_udu\right)C_{t,n},
\end{align*}
where
\[
l_u=\int_0^1(1-\rho)\lambda_{u,\rho}d\rho.
\]
This gives \eqref{ec}. 

The approximation \eqref{ec0} can be justified by assuming as in \cite{jly5}, that the portfolio is diversified among a large number $M$ of issuers so that 
\[
C_{t,n}=\sum_{i=1}^Mw^iC^i_{t,n},
\]
where $w^i$ is the relative weight and  $C^i_{t,n}$ is the outstanding payments of the $i$th issuer, respectively. The defaults of each issuer are assumed to follow a Cox process with intensity $\lambda$ as described in Section~\ref{sec:corp}. If, conditionally on $\lambda$, the defaults of different issuers are independent, we get
\begin{align*}
E(C_{s,n}-E[C_{s,n}\,|\,\lambda])^2 &= EE[(C_{s,n}-E[C_{s,n}\,|\,\lambda])^2\,|\,\lambda]\\
&= E\sum_{i=1}^M(w^i)^2E[(C^i_{s,n}-E[C^i_{s,n}\,|\,\lambda])^2\,|\,\lambda]\\
&\le E\sum_{i=1}^M(w^iC^i_{t,n})^2,
\end{align*}
where the inequality holds since since both $C^i_{s,n}$ and $E[C^i_{s,n}\,|\,\lambda]$ are bounded between zero and $C^i_{t,n}$. Assuming as in \cite{jly5}, that $C^i_{t,n}$ are uniformly bounded in $i$ and that the portfolio is well-diversified in the sense that $\sum_{i=1}^M(w^i)^2\to 0$ as $M$ grows, we get that $C_{s,n}$ converges in mean square to $E[C_{s,n}\,|\,\lambda]$. 

\bibliographystyle{plainnat}
\bibliography{bonds}

\end{document}